\documentclass[%
 reprint,
showpacs,
 amsmath,amssymb,
 aps,
]{revtex4-1}

\usepackage{graphicx}
\usepackage{dcolumn}
\usepackage{bm}

\begin{document}

\title{Hyper-Plasmonics: hyperbolic modes of a metal-dielectric interface.}



\author{Evgenii E. Narimanov}
\affiliation{School of Electrical and Computer Engineering  and Birck Nanotechnology 
Center, \\ Purdue University, West Lafayette, IN 47907, USA
}%


\begin{abstract}
Plasmon resonance, with strong coupling of light to electrons at a metal-dielectric interface, allows light confinement and control at subwavelength scale.  It's fundamentally limited by the inherent mobility of the electrons, leading to the corresponding non-locality of the electromagnetic response.\cite{Ciraci2012,Mortensen2015}
We report that this non-locality also results in the formation of a hyperbolic layer near the metal-dielectric interface, with a strong anisotropy of its electromagnetic response.  While the resulting ``hyperbolic blockade" leads to the suppression of the conventional plasmon resonance, the hyperbolic layer also supports an entirely new class of surface waves, that offer  longer propagation distance and stronger field confinement, simultaneously. Furthermore, these ``hyper-plasmons" are not limited to the proximity of the plasmon resonance, which dramatically extends the operational bandwidth of plasmonic devices.
\end{abstract}

\maketitle


With the ultimate goal of controlling light on a subwavelength scale, 
the field of nanophotonics generally relies on two main ideas -- the plasmon resonance and
the use of hyperbolic media. In the former approach, the subwavelength confinement of the electromagnetic  field is achieved via the resonant coupling to free charge carriers in a conducting
medium,\cite{Maier-book}  while in the latter its the result of the extreme anisotropy of the material response that qualitatively changes the nature of the propagating fields.\cite{ENprbrc}  These are generally considered as
fundamentally distinct concepts, with their inherent advantages and drawbacks: e.g. plasmonic systems 
that rely on the properties of a single metal-dielectric interface are generally simpler to fabricate, but are generally limited to the proximity of the corresponding resonance frequency,\cite{Maier-book} 
 while the approach based 
on hyperbolic media offers a broad bandwidth at the expense of highly nontrivial fabrication when the 
required anisotropy is due to the nanostructuring  of the material.\cite{hyperbolic-3D} However, it is now well understood that
 the fundamental limits on the light confinement in both cases are defined by the inherent non-locality
 of the electromagnetic response in the constituent materials, due to e.g. the mobility of the free carriers
 in conducting materials,\cite{Ciraci2012, Mortensen2015} In this work, we demonstrate that electromagnetic
non-locality leads to an even deeper connection between these two seemingly different concepts of plasmon resonance and hyperbolic media: the inherent mobility of the free charge carriers in 
a plasmonic material leads to a strong dielectric anisotropy near the metal-dielectric interface, where the corresponding electromagnetic response becomes effectively hyperbolic.

The resulting hyperbolic layer near the metal-dielectric interface supports a new type of surface waves
that, compared to the conventional surface plasmons, offer both longer propagation distance and stronger field localization, {\it as the same time}. This behavior is not limited to the proximity of the plasmon resonance, but -- in agreement with the generally broad bandwidth response in hyperbolic media \cite{ENprbrc} -- persists well above the
corresponding resonance frequency. Not only does this leads to a dramatic change in the resulting 
photonic density of states (by several order of magnetite) and consequently  in all associated
phenomena -- from quantum electrodynamics to nonlinear optics to near-field thermal transport, but -- by virtue of freeing plasmonics from the proximity of the corresponding plasmon resonance frequency  -- opens the field to a large class of materials that were never before considered in the context of plasmonics.

 \begin{figure}[htbp] 
   \centering
    \includegraphics[width=3.5in]{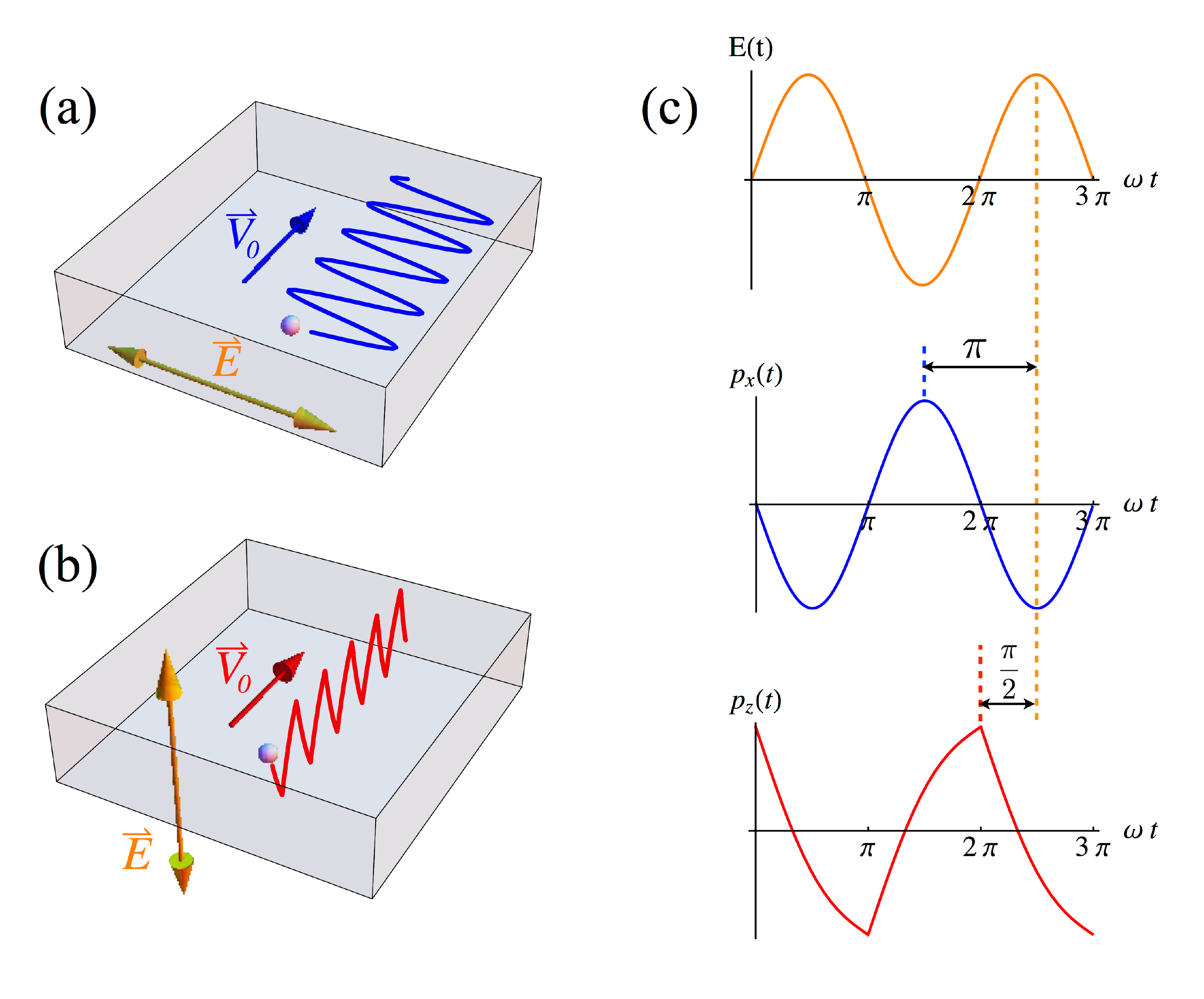}
     \caption{The classical trajectory of an electron in a thin film, with the oscillatory external field parallel (a) and perpendicular (b) to the surface. Panel (c) shows the electric field (orange) and the resulting electron dipole moment, in the tangential (blue) and normal (red) directions. Note the opposite sign
(relative phase $\pi$) of the electronic polarization in tangential direction and $\pi/2$ phase delay when the field is normal to the surface. The vector ${\bf v}_0$ in panels (a) and (b) represents the initial velocity of the electron. Note that higher harmonics visible in the red curve in (c), are removed by averaging over the actual distribution of the directions of the initial velocity.}     
     \label{fig:1}
\end{figure}

In the local approximation, the electromagnetic response of free carriers to a time-dependent electric field
depends on the corresponding frequency and the carrier scattering time, and can be defined in 
terms of the momentum transfer between the field and the free carriers. However, in a close proximity 
to a high quality metal-dielectric interface that can be considered locally flat, the electron surface reflection will reverse normal to the surface component of the momentum, while leaving its tangential
projection intact. As a result, while the specular reflection at the interface will not strongly  affect the electromagnetic response in the tangential direction, its  component that is normal to the metal surface, will be substantially altered -- leading to a strong anisotropy  in this interfacial layer. 

\begin{figure}[htbp] 
   \centering
    \includegraphics[height=6.55in]{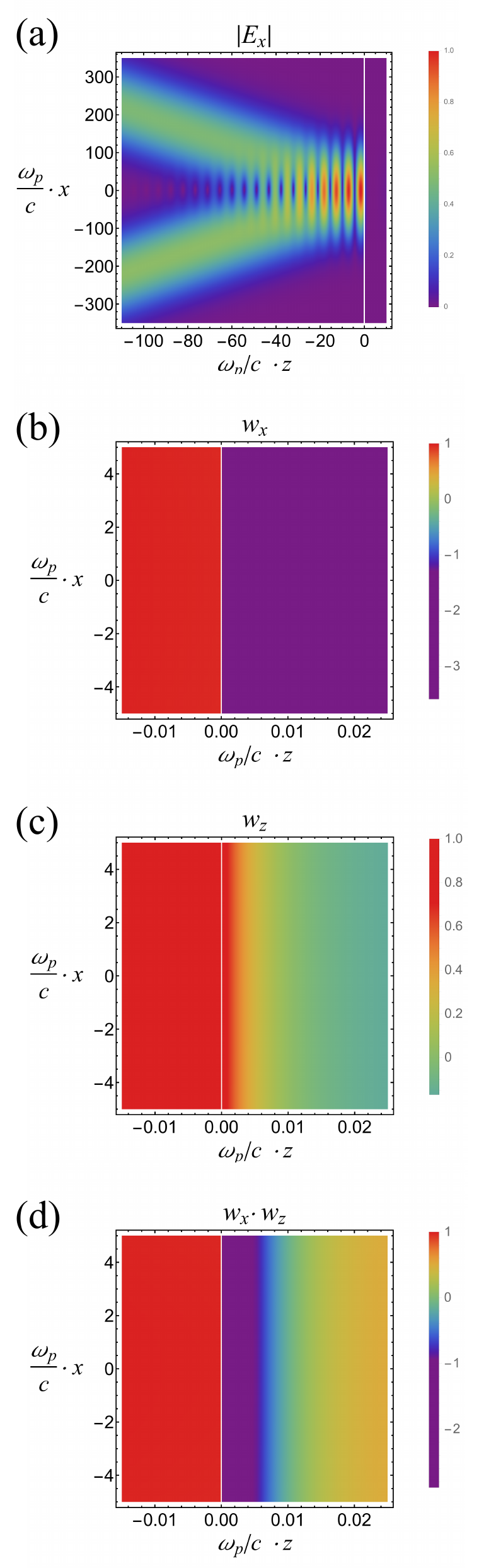}
       \caption{Gaussian beam incident on a metal-dielectric interface. Panel (a) shows the magnitude of the tangential component of the electric field. Panels (b)-(d) show the corresponding time-averaged energy density of the tangential electric field $w_x$ (panel (b)), the energy density of the normal to the interface electric field $w_z$ (panel (c)) and the product $w_x w_z$ (panel (d)). The vertical white line indicates the interface $z=0$. Note clearly visible dielectric region $z < 0$ ($w_x > 0$, $w_z > 0$), metallic region $z \gtrsim 0.01 c/\omega_p$, and the hyperbolic layer $0 < z \lesssim 0.01 c/\omega_p$. The frequency of the incident beam $\omega = 0.5 \omega_p$, the electron scattering time $\tau = 18.84 / \omega_p$, the crystal lattice permittivity of the conductor $\epsilon_\infty = 12.15$,  the permittivity of the dielectric $\epsilon_d = 10.23$, and the Fermi velocity $v_F = 0.00935$; for the plasma wavelength $\lambda_p \equiv 2 \pi c/\omega_p = 10 \ \mu{\rm m}$ these parameters correspond to the ${\rm AlInAs}/{\rm InGaAs} $ material system of Ref. \cite{ref:nmat}. Note that in this case the electron de Broglie wavelength $\lambdabar \simeq 1 \ {\rm nm}$, well below the thickness of the hyperbolic layer $(\sim 20 \ {\rm nm}$).}
          \label{fig:2}
\end{figure}

In the presence of surface roughness the free carrier reflection is no longer specular,\cite{Ziman} however the effect of the surface scattering on the momentum transfer from the free carriers to the  interface (and thus the entire sample as a whole) is 
still very different in the normal and tangential directions. As a result,  the free carrier electromagnetic response  near the conductor - dielectric interface retains its strong anisotropy.

This behavior is illustrated in Fig. \ref{fig:1}, where we consider the example of an electron that was originally moving parallel to the interface, under the parallel to the surface and perpendicular to the surface electric fields. For the field parallel to the surface, the response is similar to that  in the bulk 
medium, and the resulting contribution to the effective dipole moment ${\bf p}\left( t\right)$ and the corresponding polarization of the medium, is opposite to the field - see Fig. \ref{fig:1}(a,c), just as in the bulk of the material.
 However, when the field is driving the electron towards the surface (see Fig. \ref{fig:1}(b)), the resulting reflection from the interface reverses the sign of the normal to the interface  component of its velocity -- and the momentum initially given to the electron by the field, at the reflection is transferred to the crystal as the whole. As a result, compared to the bulk of the material, the electron response in the normal-to-the-interface direction is strongly suppressed.
As seen in Fig. \ref{fig:1}(c), the phase difference between the resulting dipole moment and the field, is now $\pi/2$, instead of the original value of $\pi$. The corresponding contribution to the permittivity is no longer negative, but imaginary, which represents the effective loss that accounts for the  transfer of the momentum from the electron to the crystal as the whole at each reflection. Without the negative contribution of the free electrons,  the real part of the permittivity in the  normal to the interface direction  is now effectively positive -- and the thin layer near the surface behaves as if it had negative permittivity parallel to the interface and positive permittivity normal to the interface. A high-quality metal-dielectric surface therefore supports a {\it hyperbolic layer}. 

\begin{figure}[htbp] 
   \centering
    \includegraphics[height=6.7in]{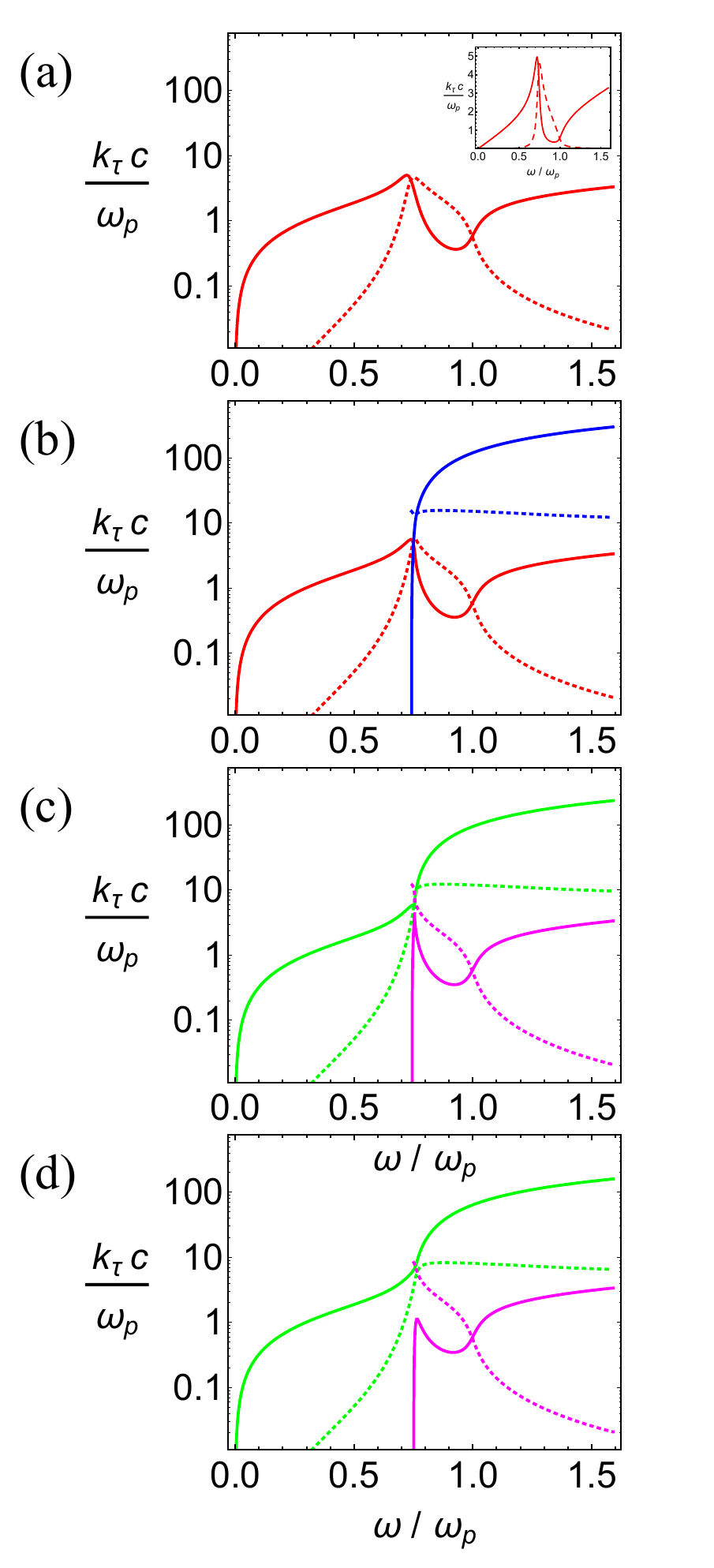}
     \caption{The dispersion of the surface waves at the metal - dielectric interface, with the in-plane momentum $k_\tau$ in units of $\omega_p / c$ and the frequency $\omega$ in units of $\omega_p$. Panel (a) corresponds to the standard result for the Drude metal, with the permittivity $\epsilon_m = \epsilon_\infty \left(1 - \omega_p^2 /\left( \omega \left(\omega + i/\tau\right) \right)  \right)$, in logarithmic scale (main panel) and linear coordinates (the inset). Panels (b) - (c) show the results for the exact solution, with the ratio of the Fermi velocity to the speed of light in vacuum, $v_F/c = 0.05$ (b), $0.0063$ (c), and $0.00935$ (d). The material parameters ($\epsilon_\infty$, $\tau$, $\epsilon_d$) are the same as in Fig. \ref{fig:2}. The red line corresponds to the conventional plasmon, blue -- to the hyperbolic mode, green  -- to the hybrid hyper-plasmon, and magenta curve -- to the suppressed resonant plasmon. With the plasma wavelength $\lambda_p  = 10 \ \mu{\rm m}$, the  doped semiconductor system ${\rm AlInAs}/{\rm InGaAs}$ corresponds to the panel (d). Note that in all cases, the wavenumber $k_\tau$ is below the Landau damping limit $\omega / v_F$.\cite{LL:physical_kinetics}}     
     \label{fig:3}
\end{figure}

The formation of the hyperbolic layer relies on  high quality of the interface that supports it. While the hyperbolic layer will adiabatically follow a smooth variation of the surface geometry, short-range surface roughness amplitude $h$ that exceeds  the characteristic scale of $v_F/\omega$, where $v_F$ is the Fermi velocity of the electrons in the metal, will suppress it. At optical frequencies, this length scale can be on the order of a few nanometers or below, and the formation of the hyperbolic layers that's predicted in the present work, is only expected in high-quality samples with sub-nanometer surface roughness. At lower frequencies however  this surface quality requirement is proportionally relaxed -- e.g. for mid-IR ``designer metals'' \cite{ref:nmat,Wasserman2015} one needs $h \lesssim 10 \ {\rm nm}$.

Note that the conventional  hydrodynamical models recently used to account for the free carrier non-locality, generally rely on the material parameters (such as e.g. the phenomenological parameter $\beta$ in Refs. \cite{Ciraci2012,Mortensen2015,BoardmanBook,Eguiluz1976,PendryHydro} ) that are taken from the {\it bulk} electromagnetic response of the conduction electrons. This approximation does not allow to describe the inherent anisotropy of the electromagnetic response in the hyperbolic layer near the metal-dielectric interface.   

Since the electromagnetic response of free charge is essentially nonlocal, 
the definition of  hyperbolic vs. dielectric vs. metallic response does not involve
local tensor of the dielectric permittivity. Instead, we rely on the general expression
of the electromagnetic energy density\cite{footnote0,LLcm} in terms of the magnetic field ${\bf B}$, electric field ${\bf E}$ and the electric displacement vector ${\bf D}$:
\begin{eqnarray}
w & = & \frac{B^2 + {\bf E}\cdot{\bf D}}{8 \pi} \equiv w^B +  w_{x,y}^E + w_z^E,
\end{eqnarray}
where $w^B$ is the magnetic field energy density,
\begin{eqnarray}
w^B & = & \frac{B^2}{8 \pi},
\end{eqnarray}
and
\begin{eqnarray}
w_{x,y}^E  =  \frac{E_x D_x + E_y D_y}{8 \pi}, \ \ \ 
w_z^E  =  \frac{E_z D_z}{8 \pi}
\end{eqnarray}
correspond to the energy densities of the electric field components  that are parallel and normal to the surface, respectively. Therefore, by definition, in a dielectric  $w_{x,y}^E > 0$ and $w_{z}^E>0$, in a metal $w_{x,y}^E < 0$ and $w_{z}^E<0$, while in a hyperbolic medium  $w_{x,y}^E$ and $w_{z}^E$ have opposite signs. For a local medium where ${\bf D} = \epsilon {\bf E}$, this reduces to the conventional definition of the dielectric, metallic and hyperbolic median in terms of $\epsilon_{x,y}$ and $\epsilon_z$.  A direct calculation or a measurement of the local electromagnetic energy density will therefore immediately uncover the specific type of the response. 

\begin{figure*}[htbp] 
   \centering
    \includegraphics[width=6.5in]{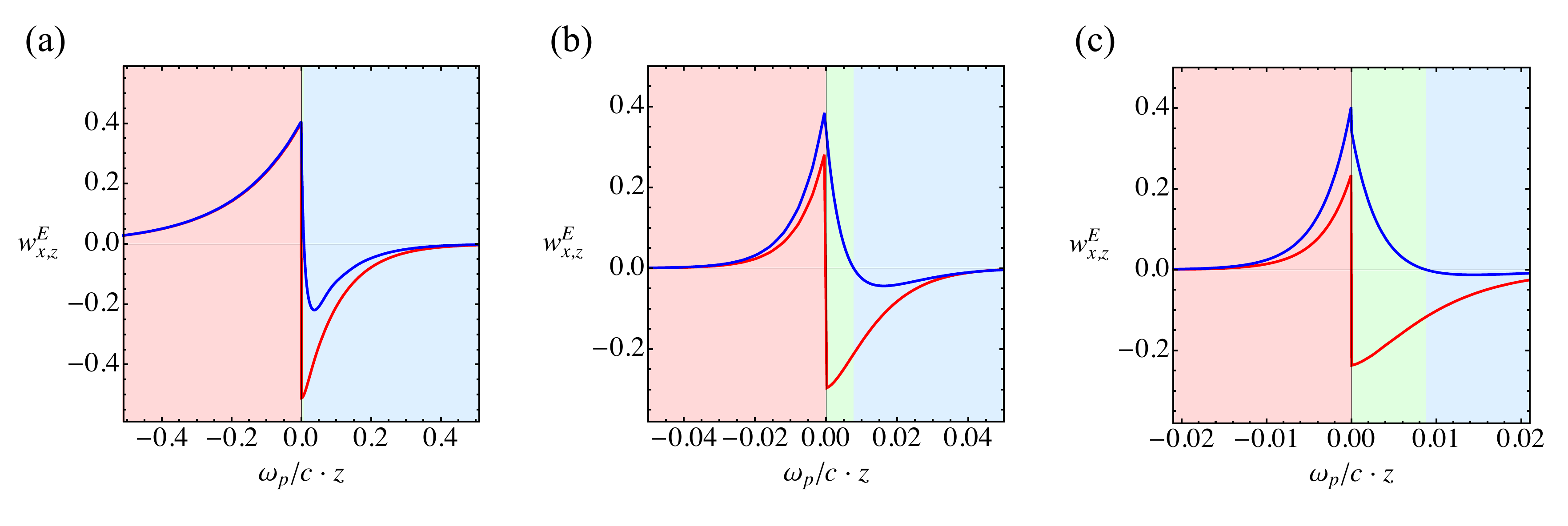}
     \caption{The hyper-plasmonic surface wave profile at the interface of isotropic dielectric with a conducting medium, and its evolution with frequency. The red and blue lines  respectively correspond to the energy densities of the tangential ($w^E_x$, blue line) and normal to the surface ($w^E_z$, red line) components of the electric field, for $\omega = 0.7 \omega_p$ (a), $\omega \simeq \omega_p $ (b),  $\omega = 1.27 \ \omega_p$ (c).The material parameters correspond to the  ${\rm AlInAs}/{\rm InGaAs} $ interface. The light-red background indicates the dielectric response  ($w_x^E > 0$ and $w_z^E > 0$), light-blue  -- to metallic response  ($w_x^E < 0$ and $w_z^E <  0$), and light-green -- to the hyperbolic layer  ($w_x^E < 0$ and $w_x^E >  0$).}
      \label{fig:4}
\end{figure*}

In Fig. \ref{fig:2} we consider a gaussian electromagnetic beam incident onto a half-infinite metal with an atomically flat  boundary at $z=0$, and calculate the actual distribution of the electromagnetic energy density that takes full account of the non-locality of the electron response in the metal.  Here, the numerical values for the plasma frequency, electron scattering time etc.  correspond to the high-quality interface of doped semiconductor GaInAs with the dielectric AlInAs, the material platform which over the last decade became the system of choose for plasmonic systems in mid-IR range. \cite{ref:nmat,Wasserman2015}
While the magnitude of the electric field (see Fig. \ref{fig:2}(a)) displays the conventional intensity pattern of the reflected wave, the plots of the local energy density (Fig. \ref{fig:2}(b)-(d)) clearly show the presence of the hyperbolic layer at $0 < z \lesssim 0.01 c/\omega_p$, where $\omega_p$ is the plasma frequency, determined from the metal's bulk response. Note that, the thickness of the hyperbolic layer in this example exceeds the electron de Broglie wavelength by more than an order of magnitude -- so 
that the formation of the hyperbolic layer can be treated within the semiclassical framework. 

The actual response to the time-dependent electromagnetic field is defined by the electronic density matrix $\rho_{{\bf p} {\bf p}'}$, governed by the Liouville - von Neumann equation \cite{ref:density-matrix}  that in the linear response regime reduces to \cite{UstinovOkulov1975,UstinovOkulov1979}
\begin{eqnarray}
\frac{\varepsilon_{\bf p} - \varepsilon_{{\bf p}'}  + \hbar \omega}{i\hbar} \ \rho_{{\bf p} {\bf p}'} + \frac{f^{(0)}_{\bf p} -f^{(0)}_{{\bf p}'} }{\varepsilon_{\bf p} - \varepsilon_{{\bf p}'}  } V^E_{{\bf p} {\bf p}'} & = & I_{{\bf p} {\bf p}'}\left[ \rho \right],
\label{eq:density_matrix}
\end{eqnarray}
where $ I_{{\bf p} {\bf p}'}\left\{ \rho \right\}$ is the collision integral that includes the contributions from both the bulk and the surface scattering of the free carriers,  $f^{(0)}_{\bf p} \equiv f_0(\varepsilon_{\bf p})$ is the equilibrium (Fermi-Dirac) distribution function, and $V_{{\bf p} {\bf p}'}$ is  the matrix element  of the spatially dependent amplitude of the electric field ${\bf E}\left({\bf r}, t\right) = {\bf E}\left({\bf r}\right) \exp\left(- i \omega t\right)$ that is given by
\begin{eqnarray}V^E_{{\bf p} {\bf p}'} & = & \int d{\bf r} \ \  {\bf j}_{{\bf p} {\bf p}'}\cdot{\bf E}\left({\bf r}\right),
\end{eqnarray}
where ${\bf j}_{{\bf p} {\bf p}'}$ is the matrix element of the charge carrier current density.\cite{ref:LL}

When the relevant ``classical'' parameters such as the mean-free path $\ell \equiv v_F \tau$ and $v_F / \omega$ are well above the free carrier de Broglie wavelength $\lambdabar$, the Wigner transformation \cite{Wigner1932,ref:LL} of the density matrix reduces \cite{KohnLuttinger,UstinovOkulov1975,UstinovOkulov1979}  Eqn. (\ref{eq:density_matrix}) to the Boltzmann equation for the charge carrier distribution function $f_{\bf p}\left({\bf r}\right)$
\begin{eqnarray}
-i\omega f_{\bf p}\left({\bf r}\right)  + {\bf v}_{\bf p} \cdot \nabla f_{\bf p}\left({\bf r}\right) +
 e {\bf E}\cdot \frac{\partial f^{(0)}_{\bf p}}{\partial {\bf p}} & = & \hat{I}\left[ f_{\bf p} \right],
 \label{eq:Boltzmann-Wigner}
\end{eqnarray}
where ${\bf v}_{\bf p} \equiv \partial\varepsilon_{\bf p}/\partial{\bf p}$ is the charge carrier group velocity for the Bloch momentum ${\bf p}$, the collision integral $\hat{I}\left[ f_{\bf p}\right]$  includes both the bulk and the surface scattering contributions, and has a highly nontrivial form. However, if the surface roughness $h$ is substantially smaller than  $v_F/\omega$ and the electron mean free path $\ell = v_F \tau$,\cite{footnote1}  the kinetic equation 
(\ref{eq:Boltzmann-Wigner}) can be expressed in the conventional form\cite{KohnLuttinger,UstinovOkulov1975,UstinovOkulov1979} 
\begin{eqnarray}
-i\omega f_{\bf p} + {\bf v}_{\bf p} \cdot \nabla f_{\bf p} +
 e {\bf E}\cdot {\bf v}_{\bf p}  \frac{\partial f_0}{\partial \varepsilon_{\bf p}} & = & - \frac{f_{\bf p}  - f_0}{\tau}
\label{eq:Boltzmann}
\end{eqnarray}
where the effective relaxation  time $\tau$ is defined by the bulk scattering, while
the effect of the surface is described by the the boundary condition on the distribution function at the interface \cite{UstinovOkulov1975,UstinovOkulov1979} -- see { Appendix A}. For a high-quality interface along one of the symmetry planes of the crystal, the latter reduces to the specular reflection boundary condition at the surface\cite{Ziman,Andreev1971,Soffer1967,Fuchs1938,Sondheimer1950,Kaner1980,ReuterSondheimer1948}
\begin{eqnarray}
f_{{\bf p}^-}\left({\bf r}_s\right) & = &f_{{\bf p}^+}\left({\bf r}_s\right),
\label{eq:ss_specular}
\end{eqnarray}
where  ${\bf p}^+$ and ${\bf p}^-$ are connected by the specular reflection condition, with equal tangential to the surface components $p_\tau^+ = p_\tau^-$, and positive and negative group velocity components in the normal to the interface direction: $\left({\bf v}_{\bf p^+}\right)_{n_s} > 0$, $\left({\bf v}_{\bf p^-}\right)_{n_s} < 0$, respectively.\cite{footnoteReflection}
 
Note that while the standard derivation \cite{BoardmanBook,Eguiluz1976} of the hydrodynamic models \cite{BoardmanBook,PendryHydro} for the electromagnetic response of free charge carriers usually follows the application of the Hamilton's principle to the Hohenberg-Kohn ground state Hamiltonian,\cite{HohenbergKohn} the hydrodynamic model can also be derived as an approximation for the solution of the kinetic equation (\ref{eq:Boltzmann-Wigner}) based on the method of moments.\cite{ChristensenThesisBook}  Such an approximation however neglects the essential anisotropy of the 
 free carrier surface scattering, and the resulting hydrodynamic approach is therefore unable to capture the formation of the hyperbolic surface layer, as well as its implications.

\begin{figure}[htbp] 
   \centering
    \includegraphics[width=3.in]{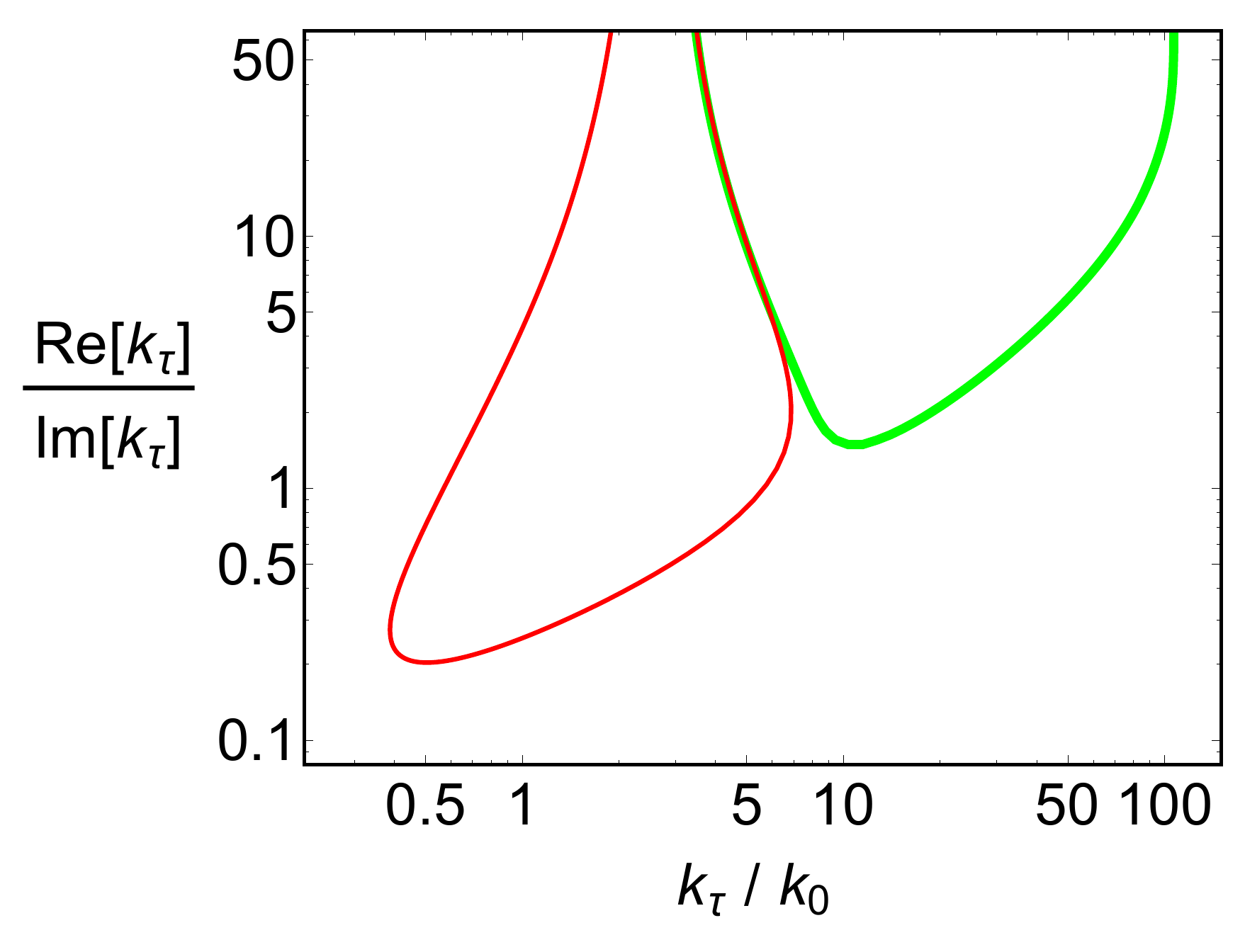}
     \caption{The ``figure-of-merit''   ${\rm Re } k_\tau / {{\rm Im} k_\tau}$ of the surface wave vs. the compression factor $k_\tau / (\omega/c)$, at the ${\rm AlInAs}/{\rm InGaAs} $ interface. Red line corresponds to the conventional plasmon and is calculated using the Drude theory (see also Fig. \ref{fig:3}(a)), while the green curve corresponds to the exact solution for the hyper-plasmon (see Fig. \ref{fig:3}(e)). }
          \label{fig:5}
\end{figure}
  
The electromagnetic field at the interface of a dielectric with a conducting medium is  defined by the self-consistent solution of the kinetic equation and the surface scattering boundary condition together with the Maxwell equations, where the electron charge and current densities  are given by
\begin{eqnarray}
\rho\left({\bf r}\right) & = & 2 \int \frac{d{\bf p}}{\left(2 \pi \hbar\right)^3} \cdot  \left(f_{\bf p}\left({\bf r}\right)  -f_0\left(\varepsilon_{\bf p} \right) \right) , \label{eqn:rho} \\
{\bf j}\left({\bf r}\right) & = & 2 \int \frac{d{\bf p}}{\left(2 \pi \hbar\right)^3} \cdot  e \  {\bf v}_{\bf p} f_{\bf p}\left({\bf r}\right),   \label{eqn:current}
\end{eqnarray}
For a high-quality planar surface, \cite{footnote2}   the corresponding mathematical problem can be reduced to the system of two coupled linear integro-differential equations (see {Appendix B}) that allows an exact analytical solution. For the the electric field in the conducting medium
we obtain
\begin{eqnarray}
{\bf E}_k\left(z > 0\right) & = & \int_{- \infty}^{\infty} \frac{dq}{2\pi} \ {\bf e}\left(k,q\right) \ \exp\left(i k x - i q z\right), \ \label{eq:e_Fourier}
\end{eqnarray}
where
\begin{eqnarray}
{\bf e}\left(k,q\right) & = & \frac{2}{D\left(k,q\right)} 
\left( 
\left. \frac{\partial E_x}{\partial z}\right|_{z = +0}
- i k
\left.  E_z\right|_{z = +0}
\right)
\nonumber \\
&  \times & \left(\epsilon_{zz}\left(k,q\right) \frac{\omega^2}{c^2} - k^2, 0,\nu_{xz}\left(k,q\right)\right), 
\end{eqnarray}
and
\begin{eqnarray}
D\left(k,q\right) & = & \left(\epsilon_{xx}\left(q\right) \frac{\omega^2}{c^2} - q^2\right)\nonumber \\
& \times & \left(\epsilon_{zz}\left(q\right) \frac{\omega^2}{c^2} - k^2 \right) - \nu^2_{xz}\left(k,q\right), 
\end{eqnarray}
with
\begin{eqnarray}
 \epsilon_{xx}\left(k,q\right) & = &\epsilon_\infty  - \frac{16 \pi i e^2 \tau}{\omega}
 \int_{v_z > 0}\frac{d{\bf p}}{\left(2 \pi \hbar\right)^3}   \frac{\partial f_0}{\partial \varepsilon_{\bf p}}
 \nonumber \\
 & \times & v_x^2 \ \frac{ 1 - i \omega \tau + i k  v_x \tau  }{ \left( 1 - i \omega \tau + i k  v_x \right)^2 +q^2 v_z^2 \tau^2 },
\end{eqnarray}
and
\begin{eqnarray}
 \epsilon_{zz}\left(k,q\right) & = &\epsilon_\infty  - \frac{16 \pi i e^2 \tau}{\omega}
 \int_{v_z > 0}\frac{d{\bf p}}{\left(2 \pi \hbar\right)^3}   \frac{\partial f_0}{\partial \varepsilon_{\bf p}}
 \nonumber \\
 & \times & v_z^2 \ \frac{  1 - i \omega \tau + i k  v_x  \tau}{ \left( 1 - i \omega \tau + i k  v_x \tau \right)^2 +q^2 v_z^2 \tau^2 },
\end{eqnarray}
and
\begin{eqnarray}
 \nu_{xz}\left(k,q\right) & = &k q -  \frac{16 \pi e^2 \tau^2 \omega q}{c^2}  
 \int_{v_z > 0}\frac{d{\bf p}}{\left(2 \pi \hbar\right)^3}  \frac{\partial f_0}{\partial \varepsilon_{\bf p}}  \nonumber \\
& \times & v_x v_z^2 \ \frac{  1 - i \omega \tau + i k  v_x  \tau}{ \left( 1 - i \omega \tau + i k  v_x \tau \right)^2 +q^2 v_z^2 \tau^2 }.
\end{eqnarray}

\begin{figure*}[htbp] 
   \centering
    \includegraphics[width=6.5in]{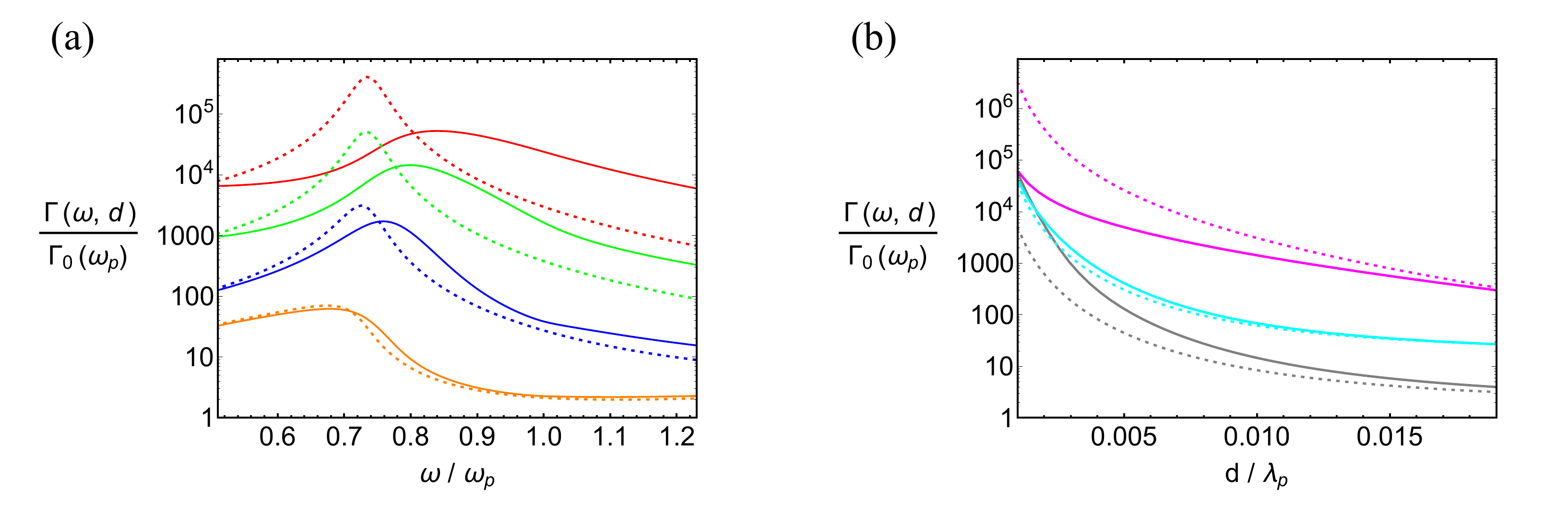}
     \caption{The spontaneous emission rate near the  dielectric - doped semiconductor interface, for the 
 ${\rm AlInAs}/{\rm InGaAs}$ system, as a function of the frequency (a) and the distance $d$ to the surface (b).
The emission rate is normalized to its value in infinite dielectric. Solid lines show the exact solution, while the corresponding dotted lines represent the results of the calculation based on the Drude theory. In panel (a), different colors corresponds to different values of the distance to the interface $\lambda_p / d = 50$ (red curves), $25$ (green), $10$ (blue) and $3$ (orange). In panel (b), different colors correspond to different frequencies, with $\omega =  0.4 \omega_p$ (cyan), $0.73 \omega_p$ (magenta) and $1.25 \omega_p$ (grey). Note that suppression of the plasmon resonance due to the hyperbolic blockage, together with and order of magnitude the enhancement of the spontaneous emission rate above the plasmon resonance frequency seen in panel (a).  }
     \label{fig:6}
\end{figure*}

For a degenerate electron gas \cite{footnote3} we reduce these expressions to
\begin{eqnarray}
\epsilon_{xx}\left(k,q\right) & = & 
\epsilon_\infty - \frac{3 \epsilon_\infty}{2} 
\frac{\omega_p^2}{\omega \left(\omega + i/\tau\right)} \left\{   
\frac{q^2 - 2 k^2}{\left(k^2 + q^2\right)^2} \frac{\left(\omega + i/\tau\right)^2}{v_F^2}  \right. \nonumber \\
& +& \left(\frac{q^2}{k^2 + q^2} + \frac{2 k^2 - q^2}{\left(k^2 + q^2\right)^2} \cdot \frac{\left(\omega + i/\tau\right)^2}{v_F^2}\right) \nonumber \\& \times & \left.  {\cal F}_0\left(\frac{v_F \sqrt{k^2 + q^2}}{\omega + i/\tau}\right)
 \right\}, \nonumber \\
\epsilon_{zz}\left(k,q\right) & = & 
\epsilon_\infty - \frac{3 \epsilon_\infty}{2} 
\frac{\omega_p^2}{\omega \left(\omega + i/\tau\right)} \left\{   
\frac{k^2 - 2 q^2}{\left(k^2 + q^2\right)^2} \frac{\left(\omega + i/\tau\right)^2}{v_F^2}  \right. \nonumber \\
& +& \left(\frac{k^2}{k^2 + q^2} + \frac{2 q^2 - k^2}{\left(k^2 + q^2\right)^2} \cdot \frac{\left(\omega + i/\tau\right)^2}{v_F^2}\right) \nonumber \\& \times & \left.  {\cal F}_0\left(\frac{v_F \sqrt{k^2 + q^2}}{\omega + i/\tau}\right)
 \right\}, \\
\nu_{xz}\left(k,q\right) & = & k q \left\{
1 +  \frac{9 \epsilon_\infty}{2}  \frac{ \omega}{\omega + i/\tau} \frac{\omega_p^2}{\left(k^2 + q^2\right) c^2 }
\right. \nonumber \\
& \times & \left.  {\cal F}_{1}\left(\frac{v_F \sqrt{k^2 + q^2}}{\omega + i/\tau}\right)
 \right\},
\end{eqnarray}
where
\begin{eqnarray}
{\cal F}_{0}\left(x\right) & = & \frac{1}{2 x}
\log\frac{1+x}{1-x},
\end{eqnarray}
and
\begin{eqnarray}
{\cal F}_{1}\left(x\right) & = & \frac{1}{x} \left\{ \frac{1}{x} + \frac{1}{2} \left( \frac{1}{3} - \frac{1}{x^2}\right)
\log\frac{1+x}{1-x}
\right\}.
\end{eqnarray}
The results presented in Fig. \ref{fig:2}, were obtained using this  solution (see Appendix C).

For a surface state at the metal-dielectric interface, matching the tangential electric field and the normal component of the electric displacement at the interface yields (see {Appendix D})
\begin{eqnarray}
\frac{1}{\pi} \int_0^\infty dq \ \frac{\epsilon_{zz}\left(k,q\right) \omega^2 / c^2  - k^2}{D\left(k,q\right)}
& = & 
- \frac{c^2}{\omega^2} \frac{\kappa_d}{\epsilon_d},
\label{eq:sw1}
\end{eqnarray}
where $\epsilon_d$ is the permittivity of the dielectric medium, and 
\begin{eqnarray}
\kappa_d & = & \sqrt{k^2 - \epsilon_d \omega^2 / c^2}
\end{eqnarray}
is the corresponding field decay rate.

Eqn.  (\ref{eq:sw1}) generally has two distinct solutions. For a sufficiently small value of the ratio of the Fermi velocity to the speed of light in vacuum, these correspond to (i) the conventional surface plasmon, and (ii) the hyperbolic wave that is primarily supported by the hyperbolic layer \cite{footnoteMultiHyper} -- see Fig. \ref{fig:3}(b). Note that the hyperbolic surface wave is only present above the cut-off frequency that is close to that of the standard surface plasmon resonance at the plant interface $\omega_{sp}$, when the bulk (Drude) metal permittivity 
\begin{eqnarray}
\epsilon_m\left(\omega\right) & = & \epsilon_\infty  \left( 1 - \frac{\omega_p^2}{\omega \left(\omega + i/\tau\right)}\right)
\label{eq:epsilon_Drude}
\end{eqnarray}
satisfies the resonance condition\cite{Maier-book}
  \begin{eqnarray}
  \epsilon_m\left(\omega_{sp} \right) = - \epsilon_d.
  \label{eq:sp}
  \end{eqnarray}

With the increase of the ratio $v_F/c$ (by e.g. increasing the doping density in a semiconductor) beyond its critical value  $(v_F/c)_*$, these two branches of the dispersion diagram undergo an avoided crossing (see {Appendix E}), so that the ``conventional'' surface plasmon continuously evolves into the hyperbolic mode
(green curve in Fig. \ref{fig:4}(c),(d)), while the plasmon resonance, with its peak in the frequency dependence of the in-plane wavenumber (and the corresponding photonic density of states), is strongly suppressed (magenta curve in Fig. \ref{fig:3}(c),(d)).  The physical origin of this suppression originates from the fact that plasmonic resonance relies on the resonant coupling between
  the electromagnetic field to the free charges in the immediate vicinity of the interface,
The formation of the hyperbolic layer with strongly anisotropic electromagnetic response, no longer
allows the resonance condition near the interface, and the conventional plasmon resonance is rapidly suppressed.

One of the main challenges in nanoplasmonics  is the inherent trade-off between the contradictory requirements of the surface plasmon propagation and field confinement.\cite{Maier-book}  In a conventional surface plasmon, an improvement of the  ``compression factor'' \cite{Marin} $k_\tau / k_0$ (that defines the field confinement) can be generally achieved only at the expense of the smaller propagation distance. This is illustrated by the red curve in Fig. \ref{fig:5}, which plots the ``figure of merit" ${\rm Re } \left[k_\tau\right] / {\rm Im}\left[ k_\tau\right]$ that represents the propagation distance in units of the plasmon's own wavelength, vs. the compression factor. However, the new ``hyper-plasmon''  surface wave that is supported by the hyperbolic layer (green curve in Fig. \ref{fig:5}) greatly exceeds the these values, for both the propagation distance and the compression factor, {\it simultaneously}. 

Due to the inherent singularity in the density of states of a hyperbolic medium,\cite{EN_prl}  the formation of the hyperbolic layer dramatically changes the photonic density of states near a high-quality metal-dielectric interface, with the resulting effect on all related phenomena -- from radiative heat transfer to quantum-electrodynamic effects to F\"orster energy transfer to nonlinear optics. As an example of this behavior, in  Fig. \ref{fig:5} we plot the spontaneous emission rate near the metal-dielectric interface, as a function of frequency (Fig. \ref{fig:5}(a)) and the distance to the interface (Fig. \ref{fig:5}(b)). Note the dramatic suppression of the conventional plasmon resonance, and the  enhancement  of the emission rate above the plasmon resonance frequency.

While the Drude theory predicts positive permittivity tensor above the plasma frequency, the inherent non-locality of the electronic response near the metal-dielectric interface dramatically modifies this simple picture. Above the plasma frequency the hyper-plasmon surface wave propagates with the in-plane wavenumber $k_\tau \gg \omega/c$, corresponding to the phase velocity $v_{\rm ph} \ll c$. For the electrons in the metal, the characteristic velocity  $v \sim v_F$ can therefore be on the order of $v_{\rm ph}$, which results in the Doppler phase shift that is comparable to the actual frequency $\omega$. As a result, even with $\omega > \omega_p$, for an electron that is propagating in the direction close to that of the surface wave, the resulting Doppler-shifted frequency
\begin{eqnarray}
\omega' & = & \omega - {\bf k} \cdot {\bf v}
\end{eqnarray}
can be well below $\omega_p$, thus increasing its negative contribution to the total permittivity
\begin{eqnarray}
\epsilon \simeq \epsilon_\infty \left(1 - \frac{\omega_p^2}{\left(\omega'\right)^2}\right).
\end{eqnarray}
Therefore, even when in the stationary frame of reference the frequency $\omega$ is well above $\omega_p$, the apparent dielectric permittivity parallel to the surface that corresponds to electromagnetic waves with large wavenumbers, is still negative. As a result, for $k \gg \omega/c$  the hyperbolic layer is still present above $\omega_p$ -- which explains the continued existence of the hyper-plasmon surface wave at higher frequencies. 

The {\it hyperbolic blockade} -- the suppression of the plasmon resonance due to the formation of the hyperbolic layer at the metal surface, caused by the inherent non-locality of the free 
electron electromagnetic response -- is the general feature of a high-quality metal-dielectric interface. However, a finite surface roughness leads to an effective averaging of the polarization anisotropy in the hyperbolic layer, and reduces the effect of the hyperbolic blockade. Quantitatively, this corresponds to the {\it short-range} roughness\cite{roughness} amplitude $h$ that exceeds the thickness of the hyperbolic layer, $\sim v_F/\omega$. Near the plasma frequency, the hyperbolic layer thickness $v_F/\omega_p$
is within a single order of magnitude from the Thomas-Fermi screening length, $v_F/\omega_p
= \left( \sqrt{3 \epsilon_\infty}/\sqrt[3]{\pi} \right)  \ R_{\rm TF}$.  In good plasmonic metals such as silver or gold, the 
hyperbolic layer thickness  can therefore be on the order of a fraction of a nanometer, and the effect of the hyperbolic blockade in all but the  highest-quality samples will be negligible. The situation however is dramatically different in other conductors, such as e.g. transparent conducting oxides\cite{TCO} or  doped semiconductors.\cite{ref:nmat,Wasserman2015} E.g. in the latter, the thickness of the hyperbolic layer is in the range between $10 \ {\rm nm}$ and  $100 \ {\rm nm}$, and exceeds both the typical roughness in high-quality MBE- or MOCVD-grown samples (generally on the order of a fraction of a nanometer) and the corresponding  electron de Broglie wavelength $\lambdabar$ by almost two orders of magnitude -- see the caption of Fig. \ref{fig:2}. Experiments on doped semiconductor materials should therefore show clear manifestations of the hyperbolic blockade, predicted in the present work.

The formation of the hyperbolic layers near the metal-dielectric interface both below {\it and} above the plasma frequency, also offers an entirely new approach for the search of new plasmonic materials. With the 
requirement for the operation in the proximity to the surface plasmon resonance frequency, the material options for nanoplasmonics remain fairly limited.\cite{TCO,Sasha_review} Although plasmonic bandwidth can be improved 
by using the metamaterial approach,\cite{meta-book} where one can design and fabricate  a metal-dielectric composite that extends the plasmonic behavior to a broader frequency range in a variety of form-factors, from planar metamaterials\cite{metamaterial-plasmons}  to core-shell plasmonic particles,\cite{core-shells}  this comes at the cost of an increased fabrication complexity,\cite{meta-book} with resulting ``hit'' in performance due to inevitable disorder at each interface.\cite{metal-losses}
  In contrast to this behavior, the hybrid ``hyper-plasmons'' introduced in the present work, offer high 
field  compression factors that are not limited to the proximity to the resonance frequency $\omega_{\rm sp}$ -- and exist well above its value (see Fig. \ref{fig:3}). To put it in the context of an actual material platform, the high-quality doped semiconductors originally introduced as plasmonic materials for mid- and far-infrared frequencies,\cite{ref:nmat}  support hyper-plasmons well into the near-IR range.

In conclusion, we introduced the concept of the hyper-plasmonic surface wave, supported by hyperbolic layers near any high-quality metal-dielectric interface. We presented the theory of this effect based on first-principles approach that takes full account of the mobility of free charge carriers in plasmonic materials and the corresponding non-locality of the electromagnetic response. For a high-quality planar interface, we obtained the exact solution of the resulting system of coupled integro-differential equations.
We demonstrated that hyper-plasmonic surface waves with simultaneously high compression factors and long propagation distance can be supported by an interface of a dielectric with conducting material, well above the corresponding plasma frequency -- thus opening the field of plasmonics to many new materials, or extending the applications of existing materials in nanophotonics to shorter wavelength.

\section*{Acknowledgments}
This work was partially supported by the National Science Foundation (grant 1629276-DMR), Army Research Office (grant W911NF-14-1-0639) and Gordon and Betty Moore Foundation.

\appendix

\section{Boundary condition for the charge carriers distribution function.}

The 
 effect of the surface can be describes by the the boundary condition on the distribution function at the interface, \cite{UstinovOkulov1975,UstinovOkulov1979} which in the general case can be expressed as
\begin{eqnarray}
f_{{\bf p}^-}\left({\bf r}_s\right) & = & \int d{\bf p}^+ \ W\left({\bf p}^-,{\bf p}^+\right) \ f_{{\bf p}^+}\left({\bf r}_s\right),
\label{eq:ss_indicatrix}
\end{eqnarray}
where the coordinate ${\bf r}_s$ corresponds to the surface, the momenta ${\bf p}^+$ and ${\bf p}^-$ correspond to the electron momenta with respectively positive and negative group velocity components in the normal to the interface direction: $\left({\bf v}_{\bf p^+}\right)_{n_s} > 0$, $\left({\bf v}_{\bf p^-}\right)_{n_s} < 0$, and the surface scattering indicatrix $W\left({\bf p}^-,{\bf p}^+\right) $ can be calculated from first principles.\cite{UstinovOkulov1975,UstinovOkulov1979} 

When the characteristic surface roughness is smaller then both the mean free path $\ell \equiv v_F \tau$ and  $v_F/\omega$,  Eqn. (\ref{eq:ss_indicatrix}) can be represented in terms of the specular reflection probability\cite{Fuchs1938,Sondheimer1950}  ${\cal P}$ as 
\begin{eqnarray}
f_{{\bf p}^-}\left({\bf r}_s\right) & = &{\cal P} \  f_{{\bf p}^+}\left({\bf r}_s\right) + \left(1 + {\cal P} \right) \Phi_\varepsilon\left(\varepsilon_{{\bf p}^-} \right),
\label{eq:ss_diffuse}
\end{eqnarray}
with ${\cal P} = 1$ corresponding to the ideal interface (\ref{eq:ss_specular}) with specular reflection and ${\cal P} = 0$ for the opposite limit of diffuse (Lambertian) scattering of the free charge carriers. Here,
 the function $\Phi_\varepsilon$ is obtained from the conservation of the electron flux to and from  the boundary. The specular reflection probability ${\cal P}$ may be treated as a phenomenological parameter, or alternatively calculated quantum-mechanically from the statistical properties of the surface roughness, \cite{Ziman,UstinovOkulov1975,UstinovOkulov1979} e.g. when the surface roughness correlation length is smaller than electron be Broglie wavelength $\lambdabar$ we find\cite{Ziman}
\begin{eqnarray}
{\cal P} & = & \exp\left( - \frac{16 \pi^2 h^2}{\lambdabar^2}\right).
\label{eq:phL}
\end{eqnarray} 

For a high-quality interface along one of the symmetry planes of the crystal, Eqn.  (\ref{eq:ss_diffuse}) reduces to the specular reflection boundary condition at the surface \cite{Ziman,UstinovOkulov1975,UstinovOkulov1979}
\begin{eqnarray}
f_{{\bf p}^-}\left({\bf r}_s\right) & = &f_{{\bf p}^+}\left({\bf r}_s\right),
\label{eq:ss_specular_SM}
\end{eqnarray}
where ${\bf p}^+$ and ${\bf p}^-$ are now connected by the specular reflection condition, with equal tangential to the surface components $p_\tau^+ = p_\tau^-$.

\section{Electromagnetic field and charge carrier distribution at the metal-dielectric interface.}

The electromagnetic field, and charge and carrier densities near the metal dielectric interface are defined 
by the self-consistent solution of the system of Maxwell's equations,
\begin{eqnarray}
{\rm div}\  {\bf D} & = & 4 \pi \rho\left({\bf r}, t\right) \label{eq:divD} \\
{\rm div} \ {\bf B} & = & 0 \label{eq:divB}  \\
{\rm curl} \ {\bf E} & = & - \frac{1}{c} \frac{\partial {\bf E}}{\partial t}\label{eq:curlE}  \\
{\rm curl} \  {\bf B} & = & - \frac{4 \pi }{c} {\bf j}\left({\bf r}, t\right)  +  \frac{1}{c} \frac{\partial {\bf D}}{\partial t},
\label{eq:curlB} 
\end{eqnarray}
where the displacement field 
\begin{eqnarray}
{\bf D} & = & \epsilon {\bf E}
=
\left\{
\begin{array}{lc}
\epsilon_d \ {\bf E}, & z < 0 \\
\epsilon_\infty \  {\bf E}, & z > 0 
\end{array}
\right. ,
\end{eqnarray}
$\epsilon_d$ is permittivity of the dielectric and $\epsilon_\infty$ is  the ``background'' permittivity of the crystal lattice in the conductor, while the free charge density $ \rho\left({\bf r}, t\right)$
and the free current density $ {\bf j}\left({\bf r}, t\right)$  are defined by the charge carrier distribution function $f_{\bf p}\left({\bf r}, t\right)$ via
\begin{eqnarray}
\rho\left({\bf r},t\right) & = & 2 \int \frac{d{\bf p}}{\left(2 \pi \hbar\right)^3} \cdot  \left(f_{\bf p}\left({\bf r}\right)  -f_0\left(\varepsilon_{\bf p} \right) \right) , \label{eq:rhoSM} \\
{\bf j}\left({\bf r}, t\right) & = & 2 \int \frac{d{\bf p}}{\left(2 \pi \hbar\right)^3} \cdot  e {\bf v}_{\bf p} f_{\bf p}\left({\bf r}, t\right).   \label{eq:currentSM}
\end{eqnarray}
In the liner response regime, the charge carrier distribution function $f_{\bf p}\left({\bf r}, t\right)$ satisfies the Boltzmann kinetic equation
\begin{eqnarray}
\frac{\partial  f_{\bf p}}{\partial t}  + {\bf v}_{\bf p} \cdot \nabla f_{\bf p} +
 e {\bf E}\cdot {\bf v}_{\bf p}  \frac{\partial f_0}{\partial \varepsilon_{\bf p}} & = & - \frac{f_{\bf p}  - f_0}{\tau},
 \label{eq:Boltzmann_SM}
\end{eqnarray}
with the boundary condition at the metal-dielectric interface (see also Eqn. (\ref{eq:ss_indicatrix}) 
\begin{eqnarray}
\left.
f_{{\bf p}}
\right|_{z = 0, v_z  < 0} & = & \int_{v'_z > 0} d{\bf p}' \ 
W\left({\bf p}, {\bf p}'\right) \  
\left.
 f_{{\bf p}'}
 \right|_{z = 0}.
 \label{eq:general_bc_SM}
\end{eqnarray}
When the surface roughness is much smaller than the charge carrier de Broglie wavelength, $h \ll \lambdabar$, or if $h \simeq \lambdabar$ and surface roughness correlation length $L \gg \lambdabar$,
Eqn. (\ref{eq:general_bc_SM}) reduces to the specular reflection boundary condition (see also Eqns. (\ref{eq:ss_specular}) and  (\ref{eq:ss_specular_SM})) 
\begin{eqnarray}
\left. f\left(v_x, v_y, v_z\right)\right|_{ z = 0} & = &\left.  f\left(v_x, v_y, - v_z\right)\right|_{ z = 0} .
\label{eq:specular_bc_SM}
\end{eqnarray}
For a harmonic wave with the in-plane momentum $k$ in the $x$-direction,
\begin{eqnarray}
{\bf E}\left({\bf r}, t\right) & = & \left( E_x\left(z\right), 0, E_z\left(z\right)\right) \ \exp\left(i k x - i \omega t\right), \label{eq:E_SM} \\
{\bf B}\left({\bf r}, t\right) & = & \left(  0, B\left(z \right),0 \right) \ \exp\left(i k x - i \omega t\right), 
\label{eq:B_SM} \\
{f_{\bf p}}\left( {\bf r}, t\right) & = &f_0\left(\varepsilon\right)  + {f}\left({\bf v}, z \right) \  \exp\left(i k x - i \omega t\right),\label{eq:f_SM}
\end{eqnarray}

Note that in the harmonic representation (\ref{eq:E_SM}),(\ref{eq:B_SM}),(\ref{eq:f_SM}), Eqns. (\ref{eq:divD}),(\ref{eq:divB}) directly follow from (\ref{eq:curlE}),(\ref{eq:curlB}), and therefore do not represent independent constrains onto the electromagnetic field and the charge carrier distribution function.\cite{RamoBook}

Applying ${\rm curl}$ to (\ref{eq:curlE}), and using (\ref{eq:curlB}), (\ref{eq:currentSM}), (\ref{eq:E_SM}), (\ref{eq:f_SM}), for $z > 0$ we obtain
\begin{eqnarray}
- \frac{\partial^2 E_x}{\partial z^2} + i k \frac{\partial E_z}{\partial z} & = & \frac{4 \pi i \omega}{c^2} j_x + \epsilon_\infty \left( \frac{\omega}{c}\right)^2 E_x, \label{eq:E1_SM} \\
i k \frac{\partial E_x}{\partial z} + k^2 E_z & = & \frac{4 \pi i \omega}{c^2} j_z + \epsilon_\infty \left( \frac{\omega}{c}\right)^2 E_z, \label{eq:E2_SM}
\end{eqnarray}
where 
\begin{eqnarray}
j_{x,z} & = & 2 e \int \frac{d{\bf p}}{\left(2 \pi \hbar\right)^3}\  v_{x,z} \ f\left({\bf v}, z\right).
\label{eq:jxz_SM}
\end{eqnarray}

Substituting (\ref{eq:f_SM}) into the kinetic equation (\ref{eq:Boltzmann_SM}) and the boundary condition (\ref{eq:specular_bc_SM}), we obtain
\begin{eqnarray}
f\left({\bf v},z\right) & = & -  e\  \frac{  \theta\left(v_z\right)}{v_z} \frac{\partial f_0}{\partial\varepsilon} \int_0^\infty d\zeta \  \left( v_x E_x\left(\zeta\right) + v_z E_x\left(\zeta\right) \right) \nonumber \\
& \times & \exp\left( - \frac{\zeta + z}{v_z}\left(\frac{1}{\tau} - i\omega +  i k v_z\right) \right) \nonumber \\
& - &  e\  \frac{  \theta\left(v_z\right)}{v_z}  \frac{\partial f_0}{\partial\varepsilon} \int_0^z d\zeta  \  \left( v_x E_x\left(\zeta\right) + v_z E_x\left(\zeta\right) \right) \nonumber \\
& \times & \exp\left[ \frac{\zeta - z}{v_z}\left(\frac{1}{\tau} - i\omega +  i k v_z\right) \right] \nonumber \\
& + & e \  \frac{\theta\left(- v_z\right)}{v_z} \frac{\partial f_0}{\partial\varepsilon} \int_z^\infty d\zeta \  \left( v_x E_x\left(\zeta\right) + v_z E_x\left(\zeta\right) \right) \nonumber \\
& \times & \exp\left[ \frac{\zeta - z}{v_z}\left(\frac{1}{\tau} - i\omega +  i k v_z\right) \right] .\label{eq:Boltzmann_solution_SM}
\end{eqnarray}

Following the approach of Ref. \cite{ReuterSondheimer1948}, originally developed in the context of the calculation of surface impedance of metals at microwave frequencies, we introduce the auxiliary fields
\begin{eqnarray}
{\cal E}_x\left(z\right) & = & E_x\left(\left|z\right|\right), \label{eq:newEx} 
\end{eqnarray}
and
\begin{eqnarray}
{\cal E}_z\left(z\right) & = &  E_z\left(\left|z\right|\right) {\rm sign} \left(z\right), \label{eq:newEz} 
\end{eqnarray}
that represent respectively even- and odd ``extension'' of the electric field in the conductor ($z>0$) to the entire range $-\infty < z < \infty$.

Substituting (\ref{eq:newEx}) and (\ref{eq:newEz}) together with (\ref{eq:jxz_SM}) and (\ref{eq:Boltzmann_solution_SM}) into (\ref{eq:E1_SM}) and (\ref{eq:E2_SM}), we obtain
\begin{eqnarray}
\frac{\partial^2 {\cal E}_x}{\partial z^2} & + & \epsilon_\infty \left(\frac{\omega}{c}\right)^2 {\cal E}_x  - i k \frac{\partial {\cal E}_z}{\partial z} \nonumber \\  = &  - & \frac{4 \pi i \omega}{c^2}  \int_{-\infty}^\infty  d\zeta \ K_{xx}\left(z - \zeta\right) \ {\cal E}_x\left(\zeta\right) \nonumber \\
&  - & \frac{4 \pi i \omega}{c^2}  \int_{-\infty}^\infty  d\zeta \ K_{xz}\left(z - \zeta\right) \ {\cal E}_z\left(\zeta\right), \label{eq:eqn1_SM} 
\end{eqnarray}
and
\begin{eqnarray}
 - i k \frac{\partial {\cal E}_x}{\partial z} & + & 
 \left(\epsilon_\infty \left(\frac{\omega}{c}\right)^2 - k^2 \right) \ {\cal E}_z 
 \nonumber \\  = &  - & \frac{4 \pi i \omega}{c^2}  \int_{-\infty}^\infty  d\zeta \ K_{zx}\left(z - \zeta\right) \ {\cal E}_x\left(\zeta\right) \nonumber \\
&  - & \frac{4 \pi i \omega}{c^2}  \int_{-\infty}^\infty  d\zeta \ K_{zz}\left(z - \zeta\right) \ {\cal E}_z\left(\zeta\right), \label{eq:eqn2_SM}
\end{eqnarray}
where 
\begin{eqnarray}
K_{xx}\left( u \right) & = & 2 \int_{v_z > 0} \frac{d{\bf p}}{\left(2 \pi \hbar\right)^3} \left(- \frac{\partial  f_0}{\partial \varepsilon}\right) \frac{v_x^2}{v_z} \nonumber \\
& \times &   \exp\left( 
 -  \left( 1 - i \omega \tau + i k  v_x \tau \right)  \frac{ \left| u \right| }{v_z \tau}
\right), \label{eq:kxx} 
\\
K_{xz}\left( u \right) & = & K_{zx}\left( u \right)  = 2 \  {\rm sign}\left(u\right)  \int_{v_z > 0} \frac{d{\bf p}}{\left(2 \pi \hbar\right)^3} \left(- \frac{\partial  f_0}{\partial \varepsilon}\right) v_x  \nonumber \\
& \times &  \exp\left( 
 -  \left( 1 - i \omega \tau + i k  v_x \tau \right)  \frac{ \left| u \right| }{v_z \tau}
\right), \label{eq:kxz} 
\\ 
K_{zz}\left( u \right) & = & 2 \int_{v_z > 0} \frac{d{\bf p}}{\left(2 \pi \hbar\right)^3} \left(- \frac{\partial  f_0}{\partial \varepsilon}\right) {v_z} \nonumber \\
& \times &   \exp\left( 
 -  \left( 1 - i \omega \tau + i k  v_x \tau \right)  \frac{ \left| u \right| }{v_z \tau}
\right). \label{eq:kzz} 
\end{eqnarray}

Despite its relative complexity, the system of coupled linear integro-differential equations (\ref{eq:eqn1_SM}),(\ref{eq:eqn2_SM}) only has difference kernels, and by means of the Fourier
transform
\begin{eqnarray}
e_x\left(k, q\right) & = & \int_{-\infty}^\infty dz \  {\cal E}_x \exp\left( i q z \right) \label{eq:ex_SM}, \\
e_z\left(k, q\right) & = & \int_{-\infty}^\infty dz \  {\cal E}_z \exp\left( i q z \right) \label{eq:ez_SM},
\end{eqnarray}
 can be reduced to a system of linear algebraic equations.\cite{Morse}  We therefore obtain
\begin{eqnarray}
{ e_x}\left(k,q\right) & = &
 \frac{2 \ A\left( k \right) }{D\left(k,q\right)} 
\left(\epsilon_{zz}\left(k,q\right) \frac{\omega^2}{c^2} - k^2\right), 
\label{eq:ex_kq_SM}
\\
{ e_z}\left(k,q\right) & = & \frac{2 \ A\left( k \right)  }{D\left(k,q\right)} 
\ \nu_{xz}\left(k,q\right), \label{eq:ez_kq_SM} 
\end{eqnarray}
where
\begin{eqnarray}
A\left( k \right) & = & \left. \frac{\partial E_x}{\partial z}\right|_{z = +0}
- i k
\left.  E_z\right|_{z = +0}, \label{eq:A_SM} \\
D\left(k,q\right) & = & \left(\epsilon_{xx}\left(q\right) \frac{\omega^2}{c^2} - q^2\right)\nonumber \\
& \times & \left(\epsilon_{zz}\left(q\right) \frac{\omega^2}{c^2} - k^2 \right) - \nu^2_{xz}\left(k,q\right), 
\label{Eq:D_SM}
\end{eqnarray}
and
\begin{eqnarray}
\epsilon_{xx}\left(k, q\right) & = & \epsilon_\infty - \frac{16 \pi i e^2 }{\omega} \int_{0}^\infty du
\cos\left(qu\right)
 \int_{v_z > 0}\frac{d{\bf p}}{\left(2 \pi \hbar\right)^3}   \frac{\partial f_0}{\partial \varepsilon_{\bf p}}\nonumber \\
& \times & 
 \frac{v_x^2}{v_z} \ 
  \exp\left( 
 -  \left( 1 - i \omega \tau + i k  v_x \tau \right)  \frac{ u}{v_z \tau}
\right), \label{eq:eps_xx_SM}  \\
\epsilon_{zz}\left(k, q\right) & = & \epsilon_\infty - \frac{16 \pi i e^2}{\omega} \int_{0}^\infty du
\cos\left(qu\right)
 \int_{v_z > 0}\frac{d{\bf p}}{\left(2 \pi \hbar\right)^3}   \frac{\partial f_0}{\partial \varepsilon_{\bf p}}\nonumber \\
& \times & 
{v_z} \ 
  \exp\left( 
 -  \left( 1 - i \omega \tau + i k  v_x \tau \right)  \frac{ u}{v_z \tau}
\right),  \label{eq:eps_zz_SM} 
\\
 \nu_{xz}\left(k,q\right) & = &k q -  \frac{16 \pi e^2}{\omega}  \int_{0}^\infty du \ 
\sin\left(qu\right)
 \int_{v_z > 0}\frac{d{\bf p}}{\left(2 \pi \hbar\right)^3}   \frac{\partial f_0}{\partial \varepsilon_{\bf p}}  \nonumber \\
& \times & 
{v_x} \ 
  \exp\left( 
 -  \left( 1 - i \omega \tau + i k  v_x \tau \right)  \frac{ u}{v_z \tau}
\right).
\label{eq:nu_xz_SM}
\end{eqnarray}

For $z>0$, the auxiliary field ${\cal \bf E}$ is identical with the the actual electric field ${\bf E}$, and Eqns. (\ref{eq:ex_SM}) - (\ref{eq:nu_xz_SM}) therefore  offer the exact analytical solution for the electric field in the metal:
\begin{eqnarray}
{\bf E}\left(z>0\right)  & = & \int_{-\infty}^\infty \frac{dq}{2 \pi} \ {\bf e}\left(k,q\right) \  \exp\left( - i q z\right).
\label{eq:e_Fourier_SM}
\end{eqnarray}

The amplitude $A(k)$ in Eqn. (\ref{eq:A_SM}) is defined by the values of the normal component of the electrical field $\left.  E_z\right|_{z = +0}$ and the  normal derivative of the tangential electric field 
 $ \left. {\partial E_x}/{\partial z}\right|_{z = +0}$ at the boundary. These magnitudes depend of the electric field in the dielectric ($z < 0$), and are obtained from the continuity of the tangential components of the electrical field and the normal components of the displacement vector
 \begin{eqnarray}
 \left. E_x\right|_{z =  -0} & = &  \left. E_x\right|_{z= + 0}, \label{eq:E_bc_SM} \\
 \epsilon_d  \left. E_z\right|_{z =  -0} & = &  \epsilon_\infty \left. E_z\right|_{z= + 0},
 \label{eq:D_bc_SM} 
 \end{eqnarray}
 where $\epsilon_d$ is the permittivity of the dielectric.

Finally, the $u$-integration in Eqns. (\ref{eq:eps_xx_SM}),(\ref{eq:eps_zz_SM}),(\ref{eq:nu_xz_SM}) 
can be performed analytically, which yields 
\begin{eqnarray}
 \epsilon_{xx}\left(k,q\right) & = &\epsilon_\infty  - \frac{16 \pi i e^2 \tau}{\omega}
 \int_{v_z > 0}\frac{d{\bf p}}{\left(2 \pi \hbar\right)^3}   \frac{\partial f_0}{\partial \varepsilon_{\bf p}}
 \nonumber \\
 & \times & v_x^2 \ \frac{ 1 - i \omega \tau + i k  v_x \tau  }{ \left( 1 - i \omega \tau + i k  v_x \right)^2 +q^2 v_z^2 \tau^2 },
\\
 \epsilon_{zz}\left(k,q\right) & = &\epsilon_\infty  - \frac{16 \pi i e^2 \tau}{\omega}
 \int_{v_z > 0}\frac{d{\bf p}}{\left(2 \pi \hbar\right)^3}   \frac{\partial f_0}{\partial \varepsilon_{\bf p}}
 \nonumber \\
 & \times & v_z^2 \ \frac{  1 - i \omega \tau + i k  v_x  \tau}{ \left( 1 - i \omega \tau + i k  v_x \tau \right)^2 +q^2 v_z^2 \tau^2 },
\\
 \nu_{xz}\left(k,q\right) & = &k q -  \frac{16 \pi e^2 \tau^2 \omega q}{c^2}  
 \int_{v_z > 0}\frac{d{\bf p}}{\left(2 \pi \hbar\right)^3}  \frac{\partial f_0}{\partial \varepsilon_{\bf p}}  \nonumber \\
& \times & v_x v_z^2 \ \frac{  1 - i \omega \tau + i k  v_x  \tau}{ \left( 1 - i \omega \tau + i k  v_x \tau \right)^2 +q^2 v_z^2 \tau^2 }. \ \ \ 
\end{eqnarray}

\section{The reflection amplitude at the planar metal-dielectric boundary.}

For a given in-plane momentum $k$, the  electric electric field in the dielectric ($z<0$) with the permittivity $\epsilon_d$ can be expressed as
\begin{eqnarray}
{\bf E}\left({\bf r}, t\right) & = & E_+  \left(1, 0, - \frac{k  }{\sqrt{\epsilon_d \left(\omega/c\right)^2 - k^2}} \right) \nonumber \\
& \times & \exp\left(i k x + i \sqrt{\epsilon_d \left(\omega/c\right)^2 - k^2} \  z - i \omega t\right) \nonumber \\
& + & E_-  \left(1, 0,  \frac{k  }{\sqrt{\epsilon_d \left(\omega/c\right)^2 - k^2}} \right) \nonumber \\
& \times & \exp\left(i k x - i \sqrt{\epsilon_d \left(\omega/c\right)^2 - k^2} \  z - i \omega t\right), \ \ \ \ \ \ \ 
\label{eq:e_field_SM}
\end{eqnarray}
leading to the corresponding magnetic field
\begin{eqnarray}
{\bf B}\left({\bf r}, t\right) & = & \frac{c}{i \omega} \  {\rm curl} \ {\bf E} \nonumber \\ 
& = & \hat{\bf y} \left[E_+ \exp\left(i \sqrt{\epsilon_d \left(\omega/c\right)^2 - k^2} z\right) \right. 
\nonumber \\
&   &\ \ \ \ -  \left. 
E_+ \exp\left(i \sqrt{\epsilon_d \left(\omega/c\right)^2 - k^2} z\right) \right] \nonumber \\
& \times & \frac{\epsilon_d\ \omega/c}{\sqrt{\epsilon_d \left(\omega/c\right)^2 - k^2}}
\exp\left( i k x - i \omega t \right).
\end{eqnarray}
Therefore the electromagnetic wave impedance\cite{RamoBook,Schelkunoff1938} in the $z = -0$ plane 
\begin{eqnarray}
\left. Z  \right|_{z = - 0} & \equiv & \left. \frac{E_x}{B_y}\right|_{z = +0} 
 = \frac{ r+1}{r-1}
\  \frac{\sqrt{\epsilon_d \left(\omega/c\right)^2 - k^2}}{\epsilon_d\ \omega/c}, \ \ \ 
\label{eq:Z_minus_SM}
\end{eqnarray}
where  the reflection coefficient
\begin{eqnarray}
r \equiv \frac{E_+}{E_-}.
\end{eqnarray}

On the other hand, from Eqns. (\ref{eq:ex_kq_SM})-(\ref{eq:e_Fourier_SM}) the tangential electric field at the metal side of the interface
\begin{eqnarray}
\left. E_x\right|_{z = +0} & = & \left( \left. \frac{\partial E_x}{\partial z}\right|_{z = +0}
- i k
\left.  E_z\right|_{z = +0} \right) \nonumber \\
& \times & \frac{1}{\pi} \int_{-\infty}^\infty dq \ 
\frac{\epsilon_{zz}\left(k, q\right) \left(\omega/c\right)^2  - k^2}{D\left(k, q\right)},
\end{eqnarray}
while the magnetic field
\begin{eqnarray}
\left. {\bf B}\right|_{z = +0} & = & \left. \frac{c}{i \omega} \  {\rm curl} \ {\bf E} \ \right|_{z = +0} \nonumber \\ 
& = & \frac{c}{i \omega} \ \hat{\bf y} \left( \left. \frac{\partial E_x}{\partial z}\right|_{z = +0}
- i k
\left.  E_z\right|_{z = +0} \right), \label{eq:B_plus_SM}
\end{eqnarray}
so that the corresponding wave impedance in the $z = +0$ plane 
\begin{eqnarray}
\left. Z  \right|_{z = + 0} & \equiv & \left. \frac{E_x}{B_y}\right|_{z = +0} \nonumber \\
& = & \frac{ i \omega}{\pi c}\int_{-\infty}^\infty dq \ 
\frac{\epsilon_{zz}\left(k, q\right) \left(\omega/c\right)^2  - k^2}{D\left(k, q\right)}.
\label{eq:Z_plus_SM}
\end{eqnarray}

From Eqns. (\ref{eq:Z_minus_SM}) and (\ref{eq:Z_plus_SM}) for the reflection coefficient $r$ we therefore obtain
\begin{eqnarray}
r & = & - 1 + 2 \left\{ 1 + i\  \frac{\epsilon_d \left(\omega/c\right)^2 }{\sqrt{\epsilon_d \left(\omega/c\right)^2 - k^2}}  \right. \nonumber \\
& \times & \left. \frac{1}{\pi}  \int_{-\infty}^\infty dq \ 
\frac{\epsilon_{zz}\left(k, q\right) \left(\omega/c\right)^2  - k^2}{D\left(k, q\right)} \right\}^{-1}.
\end{eqnarray}

\begin{figure*}[htbp] 
  \centering
    \includegraphics[width=6.5in]{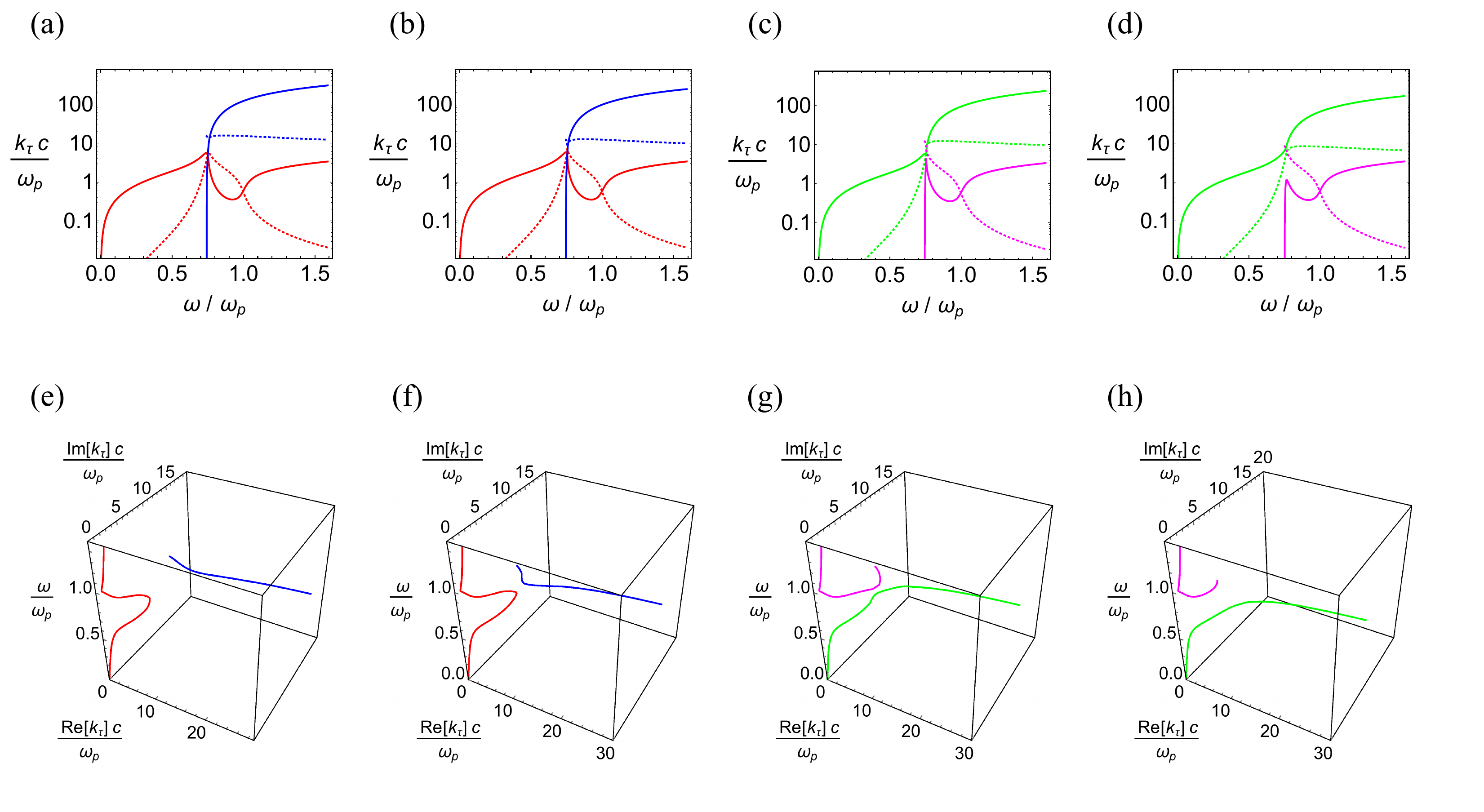}
    \caption{ The evolution of the ``crossing'' of the plasmonic (red line)  and the hyperbolic mode (blue), into the hybrid hyper-plasmonic (green line) and ``residual" (magenta) modes,   with the increase of $v_F/c$.  Panels (a,e):  $v_F/c =0.005$, panels (b,f):  $v_F/c =0.0062$, panel (c,g):  $v_F/c =0.0063$, panels (d,h):  $v_F/c =0.00935$. Other material parameters  ($\epsilon_\infty$, $\tau$, $\epsilon_d$) correspond to the 
 semiconductor system ${\rm AlInAs}/{\rm InGaAs}$, and are the same as in Fig. \ref{fig:2}.
  With the plasma wavelength $\lambda_p  = 10 \ \mu{\rm m}$, the  doped semiconductor system ${\rm AlInAs}/{\rm InGaAs}$ corresponds to the panels (d,h). 
}
\label{fig:S}
\end{figure*}

\section{Surface waves at the metal-dielectric interface.}

For a surface wave at the metal-dielectric interface with the in-plane momentum $k > \sqrt{\epsilon_d} \omega/c$, the electric field in the dielectric half-space $z<0$ is given by
\begin{eqnarray}
{\bf E}\left({\bf r}, t\right) & = & 
E_0 \left(1, 0, - \frac{i k  }{\sqrt{k^2 - \epsilon_d \left(\omega/c\right)^2}} \right) \nonumber \\
& \times & \exp\left(i k x + \sqrt{k^2 - \epsilon_d \left(\omega/c\right)^2 } \  z - i \omega t\right), \ \ \ \ \ \ \ 
\label{eq:e_field_sw_SM}
\end{eqnarray}
while the corresponding magnetic field
\begin{eqnarray}
{\bf B}\left({\bf r}, t\right) & = &\hat{\bf y} \ E_0 \  \frac{i \epsilon_d\ \omega/c}{\sqrt{k^2 - \epsilon_d \left(\omega/c\right)^2 }} \nonumber \\
& \times   & \exp\left(i k x + \sqrt{k^2 - \epsilon_d \left(\omega/c\right)^2 } \  z - i \omega t\right).
\ \ \ \ \ \ \ 
\end{eqnarray}
The wave impedance at $z = -0$ is therefore given by
\begin{eqnarray}
\left. Z \right|_{ z = - 0 } & \equiv & \left. \frac{E_x}{B_y}\right|_{z = -0}  
=  \frac{\sqrt{k^2 - \epsilon_d \left(\omega/c\right)^2 }}{i \ \epsilon_d\ \omega/c}.
\label{eq:Z_sw_SM}
\end{eqnarray}
From Eqns. (\ref{eq:Z_plus_SM}) and (\ref{eq:Z_sw_SM}) 
\begin{eqnarray}
\frac{1}{\pi }\int_{-\infty}^\infty dq \ 
\frac{\epsilon_{zz}\left(k, q\right) \left(\omega/c\right)^2  - k^2}{D\left(k, q\right)} \nonumber \\
 =  -  \frac{\epsilon_d \left( \omega/c \right)^2}{\sqrt{k^2 - \epsilon_d \left(\omega/c\right)^2 }},
\end{eqnarray}
which defines the dispersion law of the surface wave $\omega\left(k\right)$.

\section{``Crossing'' to ``Avoided Crossing'' crossover}

The dispersion equation for the surface modes at the conductor-dielectric interface, Eqn.  (\ref{eq:sw1}) generally has two distinct solutions. For a sufficiently small value of the ratio of the Fermi velocity to the speed of light in vacuum, these correspond to the conventional surface plasmon (red curve in Fig. \ref{fig:S} (a),(b) and (e),(f)), and the hyperbolic wave that is primarily supported by the hyperbolic layer (blue curve in see Fig. \ref{fig:S} (a),(b) and (e),(f)).  In this regime, there is  a large difference in the lifetimes of the ``plasmonic" and the ``hyperbolic" surface waves, so the  seemingly un-avoided crossing in the plot of the real parts of the wavenumber and the frequency in Fig. \ref{fig:S} (a),(b) is a direct consequence of this
behavior -- in the full phase space (see Fig. \ref{fig:S} (e),(f)) these two modes actually stay far apart from each other.

With the increase of the ratio $v_F/c$ (by e.g. increasing the doping density in a semiconductor) however, the corresponding lifetimes approach each other, and at the critical value of $v_F/c$ the ``plasmonic'' and the ``hyperbolic'' modes finally approach degeneracy and undergo an avoided crossing. Fig. \ref{fig:S} (c),(g) corresponds to the value of $v_F/c$ just above this critical point. From now on, with an increase of the frequency, the ``conventional'' surface plasmon continuously evolves into the hyperbolic mode -- see the evolution of the magenta curve in Fig. \ref{fig:S} (c),(d) and (g),(h). At the same time, the standard plasmonic resonance, which generally manifests itself by the peak in the frequency dependence of the in-plane wavenumber (and the corresponding photonic density of states), is strongly suppressed -- see the behavior of the magenta curve in Fig. \ref{fig:S}(c),(d) and note the use of the logarithmic scale.



\begin{thebibliography}{10}

\bibitem{Ciraci2012}
C. Ciraci, R. T. Hill, J. J. Mock, Y. Urzhumov,  A. I. Fern\'andez-Dominguez,  S. A. Maier,  J. B. Pendry,  A. Chilkoti,  and D. R. Smith, ``Probing the Ultimate Limits of Plasmonic Enhancement,''  Science
{\bf 337} (6098), 1072 - 1074  (2012).

\bibitem{Mortensen2015}
G. Toscano, J. Straubel, A. Kwiatkowski, C. Rockstuhl, F. Evers, H. Xu, N. A. Mortensen, M. Wubs, ``Resonance shifts and spill-out effects in self-consistent hydrodynamic nanoplasmonics,'' 
Nature communications {\bf 6}, 7132 (2015).

\bibitem{Maier-book}
S. A. Maier, ``Plasmonics: Fundamentals and Applications,'' (Springer; 1st edition, 2007).

\bibitem{ENprbrc} 
V. A. Podolskuiy and E. E. Narimanov, 
``Strongly anisotropic waveguide as a nonmagnetic left-handed system,'' 
Phys. Rev. B {\bf 71}, 201101(R) (2005).

\bibitem{hyperbolic-3D}
S. S. Kruk, Z. J. Wong, E. Pshenay-Severin, K. O'Brien, D. N. Neshev, Yu. S. Kivshar and X. Zhang,
``Magnetic hyperbolic optical metamaterials,'' Nature Communications {\bf 7},  11329 (2016).

\bibitem{Ziman}
J. Ziman, ``Electrons and Phonons: The Theory of Transport Phenomena in Solids ,'' (Oxford University Press; reprint edition, 2001)

\bibitem{ref:nmat}
A.~J.~Hoffman,  L.~V.~Alekseyev, S.~S.~Howard, K.~J.~Franz, D.~Wasserman, V.~A.~Podolskiy, E.~E.~Narimanov, D.~L.~Sivco and C.~Gmachl, ``Negative refraction in semiconductor metamaterials," Nature Materials {\bf 6}, 948 (2007).

\bibitem{Wasserman2015} 
Y. Zhong, S. Malagari, T. Hamilton, and D. Wasserman, ``Mid-Infrared Plasmonic Materials,''  J. Nanophotonics {\bf 9}(1), 093791 (2015).

\bibitem{BoardmanBook}
A. D. Boardman, ``Hydrodynamic theory of plasmon-polaritons on plane surfaces,'' in {\it Electromagnetic Surface Modes}, (Wiley, 1982).  

\bibitem{Eguiluz1976} 
A. Eguiluz, J. J. Quinn, ``Hydrodynamics model for surface plasmons in metals and degenerate semiconductors,'' Phys. Rev. B {\bf 14}, 1347 (1976).

\bibitem{PendryHydro} 
Y. Luo, A. I. Fernandez-Dominguez, A. Wiener, S. A. Maier, and J. B. Pendry,
``Surface Plasmons and Nonlocality: A Simple Model,''
Phys. Rev. Lett. {\bf 111}, 093901 (2013).

\bibitem{footnote0}
As we're primarily interested in optical frequencies,  the metal itself 
can be considered as a non-magnetic medium.\cite{LLcm}  However, the extension of our approach to the 
case of magnetic materials, with the energy density written in terms of the scalar product ${\bf B}\cdot {\bf H}$,\cite{LLcm}  is straightforward.

\bibitem{LLcm}
L. D. Landau, L. P. Pitaevskii, and E.M. Lifshitz, ``Electrodynamics of Continuous Media,'' 
(Butterworth-Heinemann; 2nd edition, 1984).

\bibitem{ref:density-matrix}
 H. Breuer, F. Petruccione,  ``The theory of open quantum systems,''  (Oxford University Press; 1st edition,
 2002).
 
\bibitem{UstinovOkulov1975}
V.I. Okulov, V. V. Ustinov, ``Boundary condition for the distribution function of conduction electrons scattered by a metal surface,'' Sov. Phys. JETP  {\bf 40} (3), 584 - 590  (1975).
 
\bibitem{UstinovOkulov1979}
V. I. Okulov, and V. V. Ustinov,  ``Surface Scattering of Conductivity Electrons and Kinetic Phenomena in Metals,''  Fizika Nizkikh Temperatur {\bf 5} (3),  213 - 251 (1979).

\bibitem{Wigner1932} E. Wigner,  ``On the Quantum Correction for Thermodynamic Equilibrium," Physical Review {\bf  40} (5), 749 (1932).

\bibitem{ref:LL}
L. D. Landau and  L. M. Lifshitz, ``Quantum Mechanics,'' (Butterworth-Heinemann; 3rd edition, 1981).

\bibitem{KohnLuttinger}
W. Kohn and J. M. Luttinger, ``Quantum Theory of Electrical Transport Phenomena,'' 
Phys. Rev. {\bf 108}, 590 - 611 (1957).

\bibitem{footnote1}
Here there is no restriction on the relative value of the surface roughness $h$ as compared to the de Broglie wavelength $\lambdabar$.

\bibitem{Andreev1971}
A. F. Andreev,  ``Interaction of Conducting Electrons with the Metal Surface,''  Uspekhi Fiz. Nauk {\bf  105} (1), 113 - 124 (1971). 

\bibitem{Soffer1967}
S.B. Soffer,  ``Statistical Model for the Size Effect in Electrical Conduction,''  J. Appl. Phys. {\bf  38} (4), 1710 - 1715, (1967).

\bibitem{Fuchs1938}
K. Fuchs, ``The Conductivity of Thin Metallic Films According To the Electron Theory of Metals,'' Proc Cambridge Phil. Soc. {\bf 34} (1),  100 - 108 (1938).

\bibitem{Sondheimer1950}
 E.H. Sondheimer, ``The Mean Free Path of Electrons in Metals,''  Advances in Physics {bf 50} (6),  499 - 537  (2001).

\bibitem{Kaner1980}
E. A. Kaner, A. A. Krokhin, N. M. Makarov, and V. A. Yampolskii, ``Surface absorption of electromagnetic waves in metals by random boundary inhomogeneities,'' Sov. Phys. JETP {\bf 52} (5), 938 - 944 (1980).

 \bibitem{ReuterSondheimer1948} G. E. H. Reuter and E. H. Sondheimer,
 ``The Theory of the Anomalous Skin Effect in Metals,''
Proceedings of the Royal Society of London. Series A, Mathematical and Physical Sciences {\bf  195} (1042) , 336 - 364 (1948).

\bibitem{footnoteReflection}
Note that for an anisotropic Fermi surface, ${\bf p}^-_n$ is not necessarily equal to $ - {\bf p}_n^+$. 
\cite{Ziman}

\bibitem{HohenbergKohn}
P. Hohenberg, and W. Kohn, ``Inhomogeneous electron gas,'' Phys. Rev.  {\bf 136}, 864 (1964).

\bibitem{ChristensenThesisBook}
T. Christensen, ``From Classical to Quantum Plasmonics in Three and Two Dimensions,'' (Springer, 2017).

\bibitem{footnote2}
An exact solution can be also obtained for the opposite case of diffuse scattering. While qualitatively similar, the resulting expressions show a higher degree of algebraic complexity, and will be given elsewhere.

\bibitem{footnote3}The inequality $\varepsilon_F \gg k_B T$  is generally well satisfied for plasmonic materials even at room temperature: while $k_B T \simeq 20 \ {\rm meV}$, the Fermi energy is usually in the excess of an electron-volt.

\bibitem{footnoteMultiHyper} 
Note that, while a conventional  hyperbolic waveguide supports multiple guided modes, \cite{ENprbrc} these correspond to progressively higher wavenumbers $k_\tau$. In a conducting medium however, these higher-order  modes with $k \gg v_F / \omega$ are strongly suppressed due to Landau damping. \cite{LL:physical_kinetics}

\bibitem{LL:physical_kinetics} 
 L. P. Pitaevskii, and E.M. Lifshitz, ``Physical Kinetics,'' 
(Butterworth-Heinemann; 1st edition, 1981).

\bibitem{Marin}
N. Rivera, I. Kaminer, B. Zhen, J.  D. Joannopoulos, M. Solja\u{c}i\'{c}, ``Shrinking light to allow forbidden transitions on the atomic scale,'' Science  {\bf  353} (6296), 263 - 269 (2016).

\bibitem{EN_prl}
I. I. Smolyaninov, and E. E. Narimanov, ``Metric Signature Transitions in Optical Metamaterials,'' 
Phys. Rev. Lett. 105, 067402  (2010).

\bibitem{roughness}
A slowly-varying (on the scales of the electron de Broglie wavelength and the Thomas-Fermi screening length $R_{\rm TF}$) surface roughness will simply lead to the hyperbolic layer adiabatically following its landscape, and will not suppress the hyperbolic blockade.

\bibitem{TCO}
J. Kim, G. V. Naik, N. K. Emani, U. Guler, A. Boltasseva, ``Plasmonic Resonances in Nanostructured Transparent Conducting Oxide Films,''  IEEE Journal of Selected Topics in Quantum Electronics 
{\bf 19}, 4601907, (2013).

\bibitem{Sasha_review}
G. V. Naik, V. M. Shalaev, A. Boltasseva, ``Alternative Plasmonic Materials: Beyond Gold and Silver,'' 
Advanced Materials {\bf  25} (24), 3264Ð3294 (2013). 

\bibitem{meta-book}
W. Cai and V. Shalaev, ``Optical Metamaterials: Fundamentals and Applications,'' (Springer, 1st edition, 2009).

\bibitem{metamaterial-plasmons}
Z. Liu, M. D. Thoreson, A. V. Kildishev, and V. M. Shalaev, ``Translation of nanoantenna hot spots by a metal-dielectric composite superlens,''  Appl. Phys. Lett. {\bf 95}, 033114 (2009).

\bibitem{core-shells}  
S. J. Oldenburg, R.D. Averitt, S.L. Westcott and N.J. Halas, ``Nanoengineering of optical resonances,'' Chem. Phys. Lett. {\bf 288}, 243 - 247  (1998).

\bibitem{metal-losses}
H. Reddy, U. Guler, A. V. Kildishev, A. Boltasseva, and V. M. Shalaev, ``Temperature-dependent optical properties of gold thin films,''  Optical Materials Express {\bf  6} (9), 2776 - 2802 (2016).


\bibitem{RamoBook}
S. Ramo,  J. R. Whinnery, T. Van Duzer, ``Fields and Waves in Communication Electronics,''
(John Wiley \& Sons.; 3rd edition, 1994).

\bibitem{Schelkunoff1938} S. A. Schelkunoff, ``The Impedance Concept and Its Application to Problems of Reflection, Refraction, Shielding and Power Absorption,'' Bell Syst. Tech. J. {\bf 17} (1), 17 - 48 (1938).

\bibitem{Morse} P. M. Morse and H. Feshbach, ``Methods of Theoretical Physics,'' part II (Feshbach Publishing, 2004).

\end{thebibliography}
\end{document}